\documentstyle[12pt]{article}
\begin{document}

\begin{flushright}
EDO-EP-48\\
DFPD 04/TH/19\\
September, 2004\\

hep-th/0409052\\
\end{flushright}
\vspace{20pt}

\pagestyle{empty}
\baselineskip15pt

\begin{center}
{\large\bf On the b-antighost in the Pure Spinor Quantization of
Superstrings
\vskip 1mm
}

\vspace{20mm}

Ichiro Oda
          \footnote{
          E-mail address:\ ioda@edogawa-u.ac.jp
                  }
\\

\vspace{5mm}
          Edogawa University,
          474 Komaki, Nagareyama City, Chiba 270-0198, Japan\\

\vspace{5mm}

and

\vspace{5mm}

Mario Tonin
          \footnote{
          E-mail address:\ mario.tonin@pd.infn.it
                  }
\\
\vspace{5mm}
          Dipartimento di Fisica, Universita degli Studi di Padova,\\
          Instituto Nazionale di Fisica Nucleare, Sezione di Padova,\\
          Via F. Marzolo 8, 35131 Padova, Italy\\

\end{center}


\vspace{5mm}
\begin{abstract}
Recently Berkovits has constructed a picture raised, compound 
field $b_B$ which is used to compute higher loop amplitudes in 
the pure spinor approach of superstrings. 
On the other hand, in the twisted and gauge fixed, superembedding 
approach with $n=2$ world-sheet (w.s.) supersymmetry that reproduces 
the pure spinor formulation, a field $b$ appears quite naturally 
as the current of one of the two twisted charges of the w.s. 
supersymmetry, the other being the BRST charge.
In this paper we study the relation between $b$ and $b_B$.
We shall show that $bZ$, where $Z$ is a picture raising operator,
and $b_B$ belong to the same BRST cohomological class. 
This result is of importance since it implies that the cumbersome singularity 
which is present in $b$, is in fact harmless if $b$ is combined with $Z$.

\end{abstract}

\newpage
\pagestyle{plain}
\pagenumbering{arabic}

\rm

   The pure spinor approach, developed by Berkovits in \cite{Ber1}-\cite{Ber}, 
provides a consistent quantization scheme for superstring theories with
manifest, super-Poincare covariance. Whereas untill recently only the 
prescription to compute tree level amplitudes was known, now in an important
paper \cite{Ber5} the general prescription for calculating higher genus 
amplitudes has also been proposed. Then we could say that the pure spinor approach 
provides a consistent alternative to the well-known NSR and GS formulations 
which shares the advantages of both formulations without their disadvantages.
    
\qquad To compute higher loop amplitudes in superstring theories,
a key ingredient is provided by insertions of a field $b$, 
with ghost number $-1$ which satisfies the equation 
\begin{eqnarray}
\Big\{Q, b \Big\} = T,
\label{1}
\end{eqnarray}
where  $Q$ and $T$ are the BRST charge and the stress-energy tensor,
respectively. In the pure spinor approach $Q$ and $T$ are  
given by 
\begin{eqnarray}
Q &=& \frac{1}{2 \pi i} \int dz \lambda^\alpha d_\alpha =
\oint (\lambda d), \nonumber\\
T &=& -{1\over 2} \Pi^a\Pi_a - d_\alpha \partial\theta^\alpha 
+ \omega_\alpha \partial \lambda^\alpha.
\label{2}
\end{eqnarray}
Here $\Pi^a$ and $d_\alpha$ are respectively the covariant momenta 
of the superspace coordinates $X^a$ and $ \theta^\alpha$, the ghost 
$\lambda^\alpha $ is a pure spinor, that is, a commuting spinor with 
the constraint $\lambda\Gamma^a \lambda = 0$ and $\omega_\alpha$ 
is its conjugate momentum. The action is the free field action of $ X $, 
$ \theta $, $\lambda $ and their momenta and,  as a consequence of 
the pure spinor condition, it is invariant under the local symmetry 
\begin{eqnarray}
\omega' = \omega + \Gamma^a q_a \lambda,
\label{3}
\end{eqnarray}
where $q^a $ are local gauge parameters.    
In the NSR (or GS) formulation  $b$ is the antighost of 
diffeomorphisms. On the other hand, in the pure spinor quantization 
the diffeomorphism ghosts are absent and to find a suitable $b$ is a 
non-trivial task.
     
\qquad In an attempt \cite{Matone} to understand the geometrical origin 
of the pure spinor approach it has been shown (classically and for 
the heterotic string) that the pure spinor formalism can be recovered 
as  a twisted and gauge fixed version of the superembedding formulation 
of the string with $n=2$ world-sheet (w.s.) supersymmetry. 
In this framework the existence of the pure spinor $\lambda$ 
and the absence of diffeomorphism ghosts can be understood quite naturally. 
Moreover the BRST charge of Berkovits (see also \cite{Howe}) is just one of 
the two twisted charges of the original $n=2$ supersymmetry and the field $b$ 
can be identified with the twisted current of the other charge. 
In this formulation, the $b$-ghost is of form
\begin{eqnarray}
b = {1 \over 2} (Y\Gamma_a\Pi^a d) + (\tilde\omega\partial\theta),
\label{4}
\end{eqnarray}
which indeed satisfies eq. (1). Here we have defined $Y_\alpha$ as
$Y_\alpha = {v_\alpha \over {(v\lambda)}}$ so that 
\begin{eqnarray}
(Y\lambda) = 1,
\label{5}
\end{eqnarray}
with $ v_\alpha $ being constant \cite{Oda}.
Moreover we have also defined
\begin{eqnarray}
\tilde \omega_\alpha = (\delta_\alpha ^\beta - K_\alpha \ ^\beta) \omega_\beta,
\label{6}
\end{eqnarray}
where the projector $K$ takes the form
\begin{eqnarray}
K_\alpha \ ^\beta = {1\over 2}(\Gamma_a \lambda)_\alpha (Y 
\Gamma^a)^\beta.
\label{7}
\end{eqnarray}
In this article, we adopt the following conventions: the BRST 
transformation is of form
\begin{eqnarray}
\Big\{Q, \partial\theta^\alpha\Big\} = \partial \lambda^\alpha, \ 
\Big\{Q, d_\alpha \Big\} = - \Pi^m (\Gamma_m  \lambda)_\alpha, \
\Big[Q, \Pi^m \Big] = \lambda \Gamma^m\partial \theta, \
\Big[Q, \tilde\omega_\alpha \Big] = - \tilde d_\alpha, \
\label{8}
\end{eqnarray}
where $\tilde d = (1-K)d$ (as for $\omega$). And the curly bracket
denotes the anti-commutator while the square one denotes the
commutator.
 
\qquad An expression equivalent to (4), in U(5) notations, has been given in \cite{Ber4}.  
The non-Lorentz covariance of $b$, 
in eq. (4), (due to $v_\alpha$) is not a problem since the Lorentz variation of $b$
is BRST-exact but the singular behaviour of $b$ for the configurations where 
$Y_\alpha$ diverges i.e. where $(v\lambda)= 0$, would be problematic.    
       
\qquad Here it is worthwhile to note that the strategy in \cite{Ber5} is different
where a ``picture raised'' $b_B$ field is constructed such that, instead of (1), 
it satisfies the condition \footnote{In this letter, products of field
are considered at the same point. How to treat the generic
case where T and Z are not inserted at the same point, has
been shown in \cite{Ber5}.}
\begin{eqnarray}
\Big[Q, b_B \Big] = TZ,
\label{9}
\end{eqnarray}
with Z being the ``picture raising'' operator given by 
\begin{eqnarray}
Z= {1\over 2} B_{mn}(\lambda \Gamma^{mn}d) \delta(B_{rs}N^{rs}).
\label{10}
\end{eqnarray}
Here $B_{mn}$ is a constant, antisymmetric tensor
and  the Lorentz current $N^{rs}$  is defined as 
\begin{eqnarray}
N^{rs} = {1\over 2} (\omega \Gamma^{rs}\lambda).
\label{11}
\end{eqnarray}
Then, we can easily show 
\begin{eqnarray}
\Big\{Q, Z \Big\} = 0.
\label{12}
\end{eqnarray}
Notice that $N^{rs}$ and the ghost number current 
\begin{eqnarray}
j=(\omega\lambda),
\label{13}
\end{eqnarray}
(together with $(\omega \partial \lambda)$) are the only objects involving $\omega $ 
which are invariant under the local symmetry (3). The starting point of the Berkovits' 
construction of $ b_B $ is to consider the field $\footnote{We ignore 
the normal-ordering contributions thoughout this paper.}$
\begin{eqnarray}
G^\alpha = {1\over 2} \Pi^m (\Gamma_m d)^\alpha 
- {1\over 4} N_{rs}(\Gamma^{rs}\partial \theta)^\alpha - {1 \over 4}
j \partial\theta^\alpha,
\label{14}
\end{eqnarray}
which satisfies 
\begin{eqnarray}
\Big\{Q, G^\alpha \Big\} = 
\lambda^\alpha T.
\label{15}
\end{eqnarray}
       
\qquad In this letter we shall show that eq. (4) with eq. (1) is
equivalent to eq. (14) with eq. (15) and furthermore that from eq. (4) 
one can recover the Berkovits' construction of the ``picture raised'' $b$-field
$b_B$. In particular we shall show that $ b Z$ belongs to the same BRST cohomological 
class as $b_B$.
This result leads us to two important conclusions.
One is to give support to the idea, advocated in \cite{Matone}, 
that a superembedding formulation with $n=2$ w.s. supersymmetry could be at 
the origin of the pure spinor approach. 
The other conclusion is that insertions of the simpler compound field $b$ given 
by (4)  can be used in order to compute higher genus amplitudes if $b$ is combined 
with the picture changing operator $Z$ of eq. (10) (and a trivial cocycle is added) 
since then the singular behaviour of $b$ and all the Y-dependence disappear 
from the amplitudes. 
It would be of some interest to compare the $b$-field of \cite{Ber5}
and \cite{Matone} with the $b$-field in the extended pure spinor formalism
where the pure spinor condition is removed \cite{Kazama1}-\cite{Grassi2}.
       
\qquad To verify the equivalence between eq. (4) and eq. (14),
let us consider the identity 
\begin{eqnarray}
\lambda^\alpha \omega_\beta = {1 \over 16}\delta^\alpha_\beta j 
- {1 \over 16}(\Gamma^{rs})^\alpha \ _\beta N_{rs} 
+ {1 \over 384}(\Gamma^{rspq})^{\alpha} \ _\beta (\omega\Gamma_{rspq}\lambda).
\label{16}
\end{eqnarray}
If $\tilde\omega$ is rewritten as $ \tilde\omega_\alpha = Y_\beta 
\lambda^\beta \omega_\alpha - {1\over 2} (Y\Gamma^a \omega)(\lambda
\Gamma_a)_\alpha $,
then one obtains from eq. (16) 
\begin{eqnarray}
\tilde\omega_\alpha =
Y_\beta \Big(-{1 \over 4}j \delta^\beta_\alpha 
- {1\over 4} N^{rs}(\Gamma_{rs})^{\beta} \ _{\alpha} \Big),
\label{17}
\end{eqnarray}
This result is not surprising since $\tilde \omega$ is invariant 
under the local symmetry eq. (3). With eq. (17), eq. (4)
becomes 
\begin{eqnarray}
b = Y_\alpha G^\alpha,
\label{18}
\end{eqnarray}
and from eq. (1) one has
\begin{eqnarray}
Y_\alpha \Big\{Q,G^\alpha \Big\} = Y_\alpha\lambda^\alpha T,
\label{19}
\end{eqnarray}
which coincide with eqs. (14) and (15) due to the arbitrariness of $v_\alpha$.
       
\qquad  Let us recall an important consequence of eq. (15), which
was proved in \cite{Ber5}.
For that it is convenient to introduce the following definitions: a tensor 
field $X_{\alpha_1 \cdots \alpha_n} $ with $n$  spinor indices will be 
called 
$\Gamma_5$-traceless if it vanishes when saturated with 
$(\Gamma_{a_1 \cdots a_5})^{\alpha_i\alpha_{i+1}}$ between two adjacent 
indices. Moreover, a tensor field $Y^{\alpha_1 \cdots \alpha_n}$ 
will be called pure $\Gamma_5$-trace  if $Y^{\alpha_1 \cdots 
\alpha_n} 
= \sum_{i=1}^{n-1} h_{(i)}^{\alpha_1 \cdots ((\alpha_i\alpha_{i+1}))
\cdots \alpha_n} $ where $h_{(i)}^{\alpha_1 \cdots ((\alpha_i\alpha_{i+1}))
\cdots \alpha_n}$ is symmetric in the indices 
$\alpha_i, \alpha_{i+1}$ and $\Gamma^a_{\alpha_i\alpha_{i+1}}
h_{(i)}^{\alpha_1 \cdots ((\alpha_i\alpha_{i+1})) \cdots \alpha_n}= 0$.
Then in \cite{Ber5} it is shown that eq. (15) implies the existence of the
fields $ H^{\alpha\beta}$, $K^{\alpha\beta\gamma}$, 
$L^{\alpha\beta\gamma\delta}$
and $S^{\beta\gamma\delta}$, defined modulo pure $\Gamma_5$-trace terms,
in such a way that
\begin{eqnarray}
\Big[Q, H^{\alpha\beta} \Big] = (\lambda^\alpha G^\beta)+ \cdots,
\label{20}
\end{eqnarray}
\begin{eqnarray}
\Big\{ Q,K^{\alpha\beta\gamma} \Big\} = (\lambda^\alpha H^{\beta\gamma}) +
\cdots,
\label{21}
\end{eqnarray}
\begin{eqnarray}
\Big[Q,L^{\alpha\beta\gamma\delta} \Big] = (\lambda^\alpha 
K^{\beta\gamma \delta}) + \cdots,
\label{22}
\end{eqnarray}
Moreover, since
\begin{eqnarray}
\lambda^\eta L^{\alpha\beta\gamma\delta} = 0 + \cdots,
\label{23}
\end{eqnarray}
one obtains
\begin{eqnarray}
L^{\alpha\beta\gamma\delta} = \lambda^\alpha S^{\beta\gamma\delta} +
\cdots,
\label{24}
\end{eqnarray}
where the dots in equations (20)-(24) denote pure $\Gamma_5$-trace 
terms.\par
 \qquad It is also convenient to notice that from eq. (10) one has
\begin{eqnarray}
Z=\lambda^\beta Z_\beta,
\label{25}
\end{eqnarray}
and
\begin{eqnarray}
\Big\{ Q,Z_\beta \Big\} = \lambda^\gamma Z_{\gamma\beta},
\label{26}
\end{eqnarray}
\begin{eqnarray}
\Big[ Q,Z_{\beta\gamma} \Big] = \lambda^\alpha
Z_{\alpha\beta\gamma},
\label{27}
\end{eqnarray}
and
\begin{eqnarray}
\Big\{Q,Z_{\alpha\beta\gamma} \Big\} = 
\lambda^\eta Z_{\eta\alpha\beta\gamma} + \partial\lambda^\eta \Upsilon_
{\eta\alpha\beta\gamma}.
\label{28}
\end{eqnarray}
Notice that all the Z's with more than one index and $\Upsilon $ are 
$\Gamma_5$-traceless.Altough this fact can be verified easily from 
(10), it  also 
follows directly from (12) without knowing the explicit form of $Z$.
This property is important since, as we shall see, the spinor indices of the 
fields  $ H^{\alpha\beta}$, $K^{\alpha\beta\gamma}$, 
$ L^{\alpha\beta\gamma\delta}$ and $S^{\alpha\beta\gamma}$ are saturated 
with these of the $Z$'s and $\Upsilon$, and consequently the terms of 
$\Gamma_5$-trace class,  which  are left unspecified in these fields, 
do not contribute and are irrelevant. 
     
\qquad After these preliminaries, we now turn our attention to eq. (1).
With help of eqs. (12), (18) and (25), eq. (1) can be rewritten as
\begin{eqnarray}
TZ &=& \Big[Q, bZ \Big] \nonumber\\
&=& \Big[Q, \Big(Y_\alpha(G^\alpha\lambda^\beta - \lambda^\alpha 
G^\beta) + Y_\alpha\lambda^\alpha G^\beta \Big)Z_\beta \Big].
\label{29}
\end{eqnarray}
Since $G^\alpha \lambda^\beta - G^\beta\lambda^\alpha $ is obviously 
$\Gamma_5$-traceless one can use (20) so that, taking into account (26), 
eq. (29) reduces to
\begin{eqnarray}
TZ = \Big[Q, Y_\alpha(H^{\alpha\beta} - H^{\beta\alpha})\lambda^\gamma 
Z_{\gamma\beta} \Big] + \Big[Q, b_1 \Big],
\label{30}
\end{eqnarray}
where $b_1$ is defined as
\begin{eqnarray}
b_1 =  G^\beta Z_\beta.
\label{31}
\end{eqnarray}
Since we can see that the $\Gamma_5$-trace with respect to the indices
$\gamma,\beta$ in 
$(H^{\alpha\beta} - H^{\beta\alpha})\lambda^\gamma - H^{\gamma\beta}
\lambda^\alpha$ does not contribute since $Z_{\gamma\beta}$ is 
$\Gamma_5$-traceless, we have using eqs. (21) and (27)
\begin{eqnarray}
TZ &=& \Big[Q, Y_\alpha\Big((H^{\alpha\beta} - H^{\beta\alpha})\lambda^\gamma
- \lambda^\alpha H^{\gamma\beta} \Big) Z_{\gamma\beta} \Big] 
+ \Big[Q, b_1 + b_2 \Big], \nonumber\\
&=& \Big[Q, Y_\alpha (- K^{\alpha\beta\gamma} - K^{\gamma\beta\alpha} 
+ K^{\gamma\alpha\beta} ) \lambda^\eta Z_{\eta\gamma\beta} \Big] 
+ \Big[Q, b_1 + b_2 \Big],
\label{32}
\end{eqnarray}
where we have defined  
\begin{eqnarray}
b_2 = H^{\gamma\beta} Z_{\gamma\beta}.
\label{33}
\end{eqnarray}
Using (22) and (28), the same procedure can be repeated once again to get 
\begin{eqnarray}
TZ = \Lambda ^{(a)} + \Lambda^{(b)} + 
\Big[Q, b_1 + b_2 + b_3 \Big],
\label{34}
\end{eqnarray}
where we have defined
\begin{eqnarray}
b_3 = - K^{\eta\gamma\beta} Z_{\eta\gamma\beta},
\label{35}
\end{eqnarray}
and
\begin{eqnarray}
\Lambda^{(a)} = \Big[Q, Y_\alpha(L^{\alpha\eta\gamma\beta} 
+L^{\eta\gamma\alpha\beta} - L^{\eta\gamma\beta\alpha} 
- L^{\eta\alpha\gamma\beta})\lambda^\epsilon 
Z_{\epsilon\eta\gamma\beta} \Big],
\label{36}
\end{eqnarray}
\begin{eqnarray}
\Lambda^{(b)} = \Big[Q, Y_\alpha(L^{\alpha\eta\gamma\beta} 
+L^{\eta\gamma\alpha\beta} - L^{\eta\gamma\beta\alpha} 
- L^{\eta\alpha\gamma\beta}) \partial\lambda^\epsilon 
\Upsilon_{\epsilon\eta\gamma\beta} \Big].
\label{37}
\end{eqnarray}
$\Lambda^{(a)}$ can furthermore be rewritten as
\begin{eqnarray}
\Lambda^{(a)} = \Big[Q, Y_\alpha \Big((L^{\alpha\eta\gamma\beta} 
+L^{\eta\gamma\alpha\beta} - L^{\eta\gamma\beta\alpha} 
- L^{\eta\alpha\gamma\beta})\lambda^\epsilon -
L^{\epsilon\eta\gamma\beta}\lambda^\alpha \Big) 
Z_{\epsilon\eta\gamma\beta} \Big] +
\Big[Q, b_{4}^{(a)} \Big],
\label{38}
\end{eqnarray}
where we have defined
\begin{eqnarray}
b_{4}^{(a)} = L^{\epsilon\eta\gamma \beta}Z_{\epsilon\eta\gamma\beta}.
\label{39}
\end{eqnarray}
Since $Z_{\epsilon\eta\gamma\beta}$ is $\Gamma_5 $-traceless, the first 
term in (38) vanishes owing to (23).
Accordingly, $\Lambda^{(a)}$ is expressed in terms of a BRST-exact term
\begin{eqnarray}
\Lambda^{(a)} =\Big[Q, b_{4}^{(a)} \Big].
\label{40}
\end{eqnarray}
As for $\Lambda^{(b)} $ it can be written as 
\begin{eqnarray}
\Lambda^{(b)} =\Big[Q, b_{4}^{(b)} \Big],
\label{41}
\end{eqnarray}
where we have defined 
\begin{eqnarray}
b_{4}^{(b)} &=& Y_\alpha \Big[Y_\kappa\lambda^\kappa( L^{\alpha\eta\gamma\beta} 
+ L^{\eta\gamma\alpha\beta} - L^{\eta\gamma\beta\alpha} - L^{\eta\alpha\gamma\beta}) 
-  Y_\kappa L^{\kappa\eta\gamma\beta}\lambda^\alpha \Big] 
\partial\lambda^\epsilon \Upsilon_{\epsilon\eta\gamma\beta} \nonumber\\
&+& Y_\alpha Y_\kappa \lambda^\alpha L^{\kappa\eta\gamma\beta}
\partial\lambda^\epsilon \Upsilon_{\epsilon\eta\gamma\beta}.
\label{42}
\end{eqnarray}
As before the first term vanishes and then using eqs. (5) and (24), one has
\begin{eqnarray}
b_{4}^{(b)} = S^{\eta\gamma\beta}\partial\lambda^\epsilon \Upsilon_{
\epsilon\eta\gamma\beta}.
\label{43}
\end{eqnarray}

\qquad Consequently, from eqs. (34), (40) and (41), we have recovered eq. (9) 
where
\begin{eqnarray}
b_B = b_1 +b_2 +b_3 + b_{4}^{(a)} + b_{4}^{(b)}.
\label{44}
\end{eqnarray}
Note that $ b_1$, $b_2$, $b_3$, $b_{4}^{(a)}$, $b_{4}^{(b)}$ are respectively
given in eqs. (31), (33), (35), (39), (43) which are in complete agreement 
with the result of ref. \cite{Ber5}.  Also notice that our construction
shows that $bZ$ and $b_B$ belong to the same BRST cohomological class, 
as promised. A related but alternative and interesting recipe to compute one-loop
amplitudes has been recently proposed in \cite{Anguelova}. It is of interest 
to remark that if in eq. (28) one replace $Z$ with the unintegrated vertex operator 
$V= \lambda^\beta V_\beta $ that it is needed at one loop level
and then perform the same manipulations that lead from eq. (28) to eq. (43), 
one ends with eq. (5.25) of \cite{Anguelova}. 

\qquad Now a remark is in order. 
As can be seen in our above derivation, $S^{\alpha\beta\gamma}$ depends on 
$Y_\alpha$ through its dependence on $\tilde\omega$, so
one might worry that this dependence could remain even in $b_{4}^{(b)}$
from eq. (43). In that case, $b_{4}^{(b)}$ could become singular at $(v\lambda) = 0$,
thereby inducing troublesome divergences in the loop amplitudes.
However, luckily enough, this is not the case. To show that, let us notice that 
$\Big[ Q, b_{4}^{(b)}\Big]$ must be regular from eqs. (9) and (44) (noting that
$T$ and the other terms of $b_B$ are regular) and therefore is independent of 
$Y_\alpha$. 
Moreover, on general ground and modulo harmless contributions with 
only regular $j$ , $N^{rs}$, and  $ (\omega\partial\lambda)$, 
it turns out that $b_{4}^{(b)}$ is expressed by a linear combination of
the following terms:
\begin{eqnarray}
b_{4}^{(b)} &=&   
x B_{pq}(\tilde\omega \Gamma^{pq}\partial\lambda) j 
+ y B_{pq}(\tilde\omega\Gamma^{pqrs}\lambda)(\partial\lambda\Gamma_{rs}\tilde\omega) 
+ z B_{pq}(\tilde\omega\Gamma^{prst}\lambda)(\tilde\omega
\Gamma^{q}_{\quad rst}\partial\lambda)  \nonumber\\
&+& v B_{pr}(\tilde\omega\Gamma^{rs}\partial
\lambda) N_s \ ^p + w B_{pq}(\tilde\omega\Gamma^{pqrs}\partial\lambda)N_{rs},
\label{45}
\end{eqnarray}
with $x$, $y$, $z$, $v$, and $w$ being constants. 
By a repeated use of the Fierz identity and taking into account the pure
 spinor condition $\lambda \Gamma^a\lambda = 0$,
the last term of this equation reduces to a combination of the first two
terms in this equation and of the fourth one   modulo harmless 
contributions and the fourth term reduces to the first one modulo harmless 
contributions..  
Then $b_{4}^{(b)}$ reduces to a linear combination of the first 3 terms of 
this equation and regular contributions independent of $Y_\alpha$. 
The BRST variations of the these 3 terms depend on $Y_\alpha$ in an 
independent way. Indeed the Y-dependence
 of the variation of the first term is proportional to $(\lambda\Gamma^
{apq}\partial\lambda)$,  that of the second term is proportional to
$(\lambda\Gamma^{apqrs}\lambda)$ and that of the third one to 
$ (\lambda\Gamma ^{pqrst}\partial\lambda)$. Therefore                
 the coefficients of these terms must vanish separately and $b_{4}^
{(b)}$ is regular.

\newpage
\begin{flushleft}
{\bf Acknowledgements}
\end{flushleft}

We are grateful to M. Matone, P. Pasti and D. Sorokin for stimulating 
discussions, and thank N. Berkovits for useful discussions and
comments.
The first author (I.O.) would
like to thank Dipartimento di Fisica, Universita degli Studi di Padova
for its kind hospitality and his work was partially supported by
the Grant-in-Aid for Scientific Research (C) No.14540277 from 
the Japan Ministry of Education, Science and Culture.
The work of the second author (M.T.) was partially supported by the European
Community's Human Potential Programme under contract HPRN-CT-2000-00131
Quantum Spacetime and by the INTAS Research Project No.2000-254.


\end{document}